\documentclass[aps,pre,preprint,groupedaddress,showpacs]{revtex4}

\usepackage[dvips]{graphicx}
\usepackage{amsmath}

\begin{document}

\title{Experimental synchronization of circuit oscillations\\
induced by common telegraph noise}

\author{Ken Nagai$^1$ and Hiroya Nakao$^{1,2}$}

\affiliation{ $^1$Department of Physics, Graduate School of Science,
  Kyoto University, Kyoto 606-8502, Japan
  \\
  $^2$Abteilung Physikalische Chemie, Fritz-Haber-Institut der
  Max-Planck-Gesellschaft, Faradayweg 4-6, 14195 Berlin, Germany}

\date{\today}

\begin{abstract}
  Experimental realization and quantitative investigation of 
  common-noise-induced synchronization of limit-cycle oscillations
  subject to random telegraph signals are performed using an
  electronic oscillator circuit.
 Based on our previous formulation~[K.~Nagai, H.~Nakao, and Y.~Tsubo, Phys.~Rev.~E {\bf 71}, 036217 (2005)], dynamics of the circuit is described as random phase mappings between two limit
  cycles. Lyapunov exponents characterizing the degree of
  synchronization are estimated from experimentally determined phase
  maps and compared with linear damping rates of phase differences
  measured directly.  Noisy on-off intermittency of the phase difference
  as predicted by the theory is also confirmed experimentally.
\end{abstract}

\pacs{82.40.Bj,89.75.Da,43.50.+y}

\maketitle

\section{Introduction}

Synchronization of nonlinear dynamical elements is observed in many
natural systems~\cite{Kuramoto_book1984,Strogatz1994}.
For instance, in our body, heart cells synchronize with each other to
generate heartbeats, and suprachiasmatic neurons synchronize with the
24 hour daily cycle to generate circadian
rhythms~\cite{Winfree_book2001,RusakZucker1979}.
Many experimental investigations of synchronization have been
  carried out, e.g. using coupled chemical
  reactors~\cite{MarekStuchl1975,YoshimotoYoshikawaMori1993,FukudaMorimuraKai2005}.
Synchronization typically occurs due to mutual coupling or through 
entrainment to common periodic signals. Generally, external noises 
independently applied to the elements have negative effects on 
synchronization; the elements cannot synchronize under 
independent noise sources that are too extreme.

In contrast, common or correlated external noises can synchronize
uncoupled dynamical elements. Using neurons of rat neocortical slices,
Mainen and Sejnowski~\cite{MainenSejnowski1995} have shown
that reliability of
spike generation improves when a neuron receives a fluctuating input
current compared with the case of a constant input current. This
phenomenon can be considered as synchronization of uncoupled identical
dynamical elements induced by common fluctuating inputs.
The synchronizing effect of common fluctuating forcing, known in
ecology as the Moran effect, describes the synchronized population
dynamics of organisms due to correlated environmental
fluctuations~\cite{Grenfell1998}.

More explicitly, Pikovskii~\cite{Pikovskii1984} and
Jensen~\cite{Jensen2002} have theoretically investigated
synchronization of limit-cycle oscillators induced by non-periodic external
signals. Synchronization of uncoupled chaotic oscillators due to
common noisy driving has been numerically studied by Maritan and
Banavar~\cite{MaritanBanavar1994} and experimentally realized by
S\'{a}nchez et al.~\cite{Sanchez1997} using an electronic circuit.
Synchronization (or consistency) of chaotic lasers due to common
fluctuating signals has also been reported~\cite{UchidaMcAllisterRoy2004}.

For limit-cycle oscillators, general quantitative formulations of
common-noise-induced synchronization can be developed using the
phase-reduction method~\cite{Kuramoto_book1984}.
Teramae and Tanaka~\cite{Teramae2004} have proven that the synchronized
state of uncoupled limit cycle oscillators subject to a common weak
Gaussian noise is always statistically stabilized, and their theory
has been further generalized to provide global stability of phase coherent
states induced by correlated noises~\cite{NakaoAraiKawamura2007}. The
cases where limit-cycle oscillators are stimulated by a common
telegraph noise~\cite{NagaiNakaoTsubo2005} or by a common impulsive
noise~\cite{NakaoAraiNagaiTsuboKuramoto2005} have also been investigated
theoretically using random phase-map descriptions.
Recently, synchronization due to common random
impulses has been studied experimentally with an electronic circuit, and
some of the theoretical predictions have been quantitatively verified~\cite{AraiNakao2008}.

In this paper, we experimentally investigate common-noise-induced synchronization using an electronic circuit undergoing periodic oscillations. As the random signal, we use a random telegraph noise, which is the simplest example of colored non-Gaussian noises; it can easily be generated in experiments and facilitates analytical treatments. In this case, we have two limit cycles corresponding to two values of the driving signal, in contrast to the previous experiment using random impulses where the system possessed only a single limit-cycle orbit~\cite{AraiNakao2008}. 
When the switching interval of the driving signal is sufficiently
long, the circuit state is mostly on either of the limit cycles
at the moments of switching, so that its dynamics can be
described in terms of phase mappings between the two limit cycles.
We experimentally determine the phase maps of the electronic circuit
and estimate the Lyapunov exponents characterizing the degree of
synchronization from the phase maps based on our previous theory.
The Lyapunov exponents are then quantitatively compared with the
damping rates of small phase differences measured directly.
We also confirm that noisy on-off intermittency of the phase
difference, which is typically expected for random-mapping
systems~\cite{Fujisaka1985,Cenys1997,Pikovsky1992,Nakao1998,
  NagaiNakaoTsubo2005}, actually occurs in our electronic circuit.

\section{Experiments}

\subsection{Setup}

\begin{figure}[htbp]
  \includegraphics[width=0.9\linewidth]{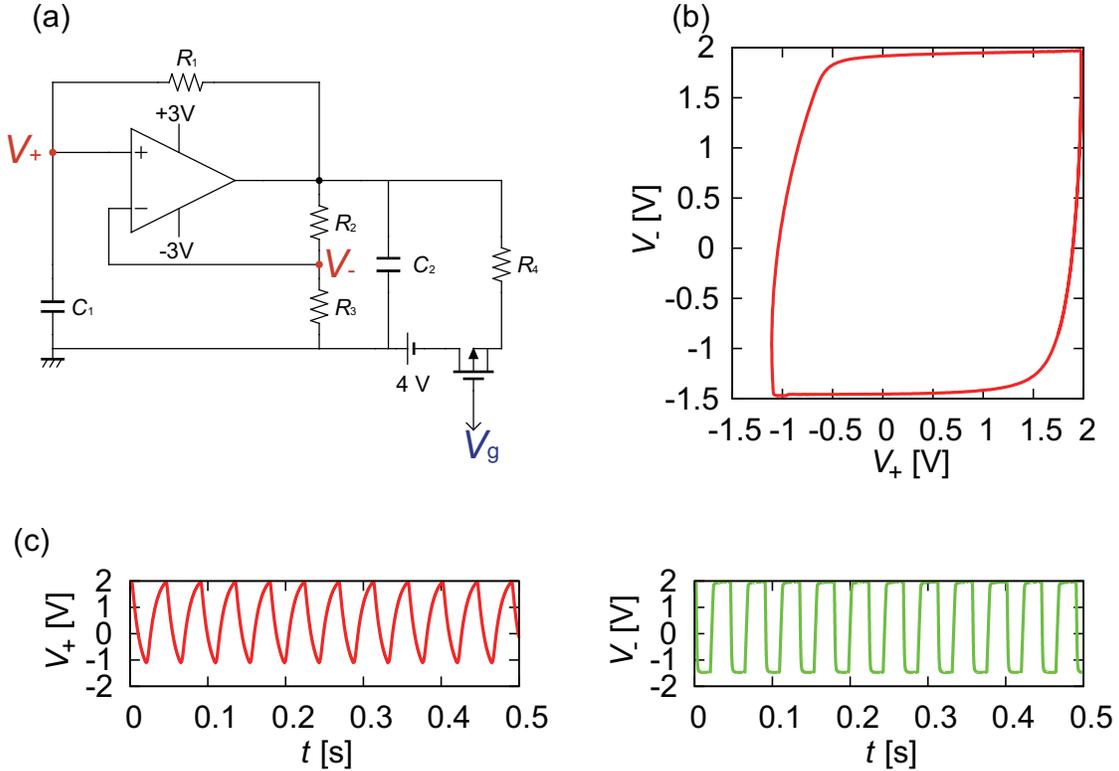}
  \caption{(Color online) Experimental setup.  (a) An electronic circuit
    used in the experiment.  (b) A limit-cycle orbit observed at
    $V_{\rm{g}}(t) \equiv -6.0\;\rm{V}$ on the ($V_{+}$, $V_{-}$)
    plane.  Average period was 0.0445~s.  (c) Time series of
    $V_{+}(t)$ (left) and $V_{-}(t)$ (right).}
 \label{circuit}
\end{figure}

The experiments were performed using an electronic circuit shown
schematically in Fig.~\ref{circuit}(a), where an LM741 was used as the
Op-Amp and the circuit parameters were set as follows:
$R_{1}=100\;{\mathrm k}\Omega \pm 5\;{\mathrm k}\Omega$,
$R_{2}=81\;{\mathrm k}\Omega\pm4.5\;{\mathrm k}\Omega$,
$R_{3}=470\;{\mathrm k}\Omega\pm 23.5\;{\mathrm k}\Omega$,
$R_{4}=100\;\Omega\pm5\;\Omega$, $C_{1}=100\;\mathrm{nF}\pm
5\;\mathrm {nF}$, and $C_{2}=10\;\mu {\mathrm F}\pm 1\;\mu
{\mathrm F}$. Voltages of positive and negative power supplies to the
Op-Amp were fixed at $\pm$3.0~V (0~V indicates the
ground voltage) using a DC power source (PMM18-2.5DU, Kikusui
Electronics Co.).
Note that we use the Op-Amp under positive feedback conditions to generate oscillations, so that the golden rule of an Op-Amp ($V_{+}=V_{-}$) does not hold in our experiments.

The source voltage of the MOSFET (2SK2201, Toshiba Co.) was fixed at
-4.0~V with another DC power source (E3630A, Hewlett Packard)
and the gate voltage $V_{\rm{g}}(t)$ was controlled by the external
signal.
Voltage traces $V_{+}(t)$ and $V_{-}(t)$ were measured from the
circuit as shown in Fig.~\ref{circuit}(a). Control of $V_{\rm{g}}(t)$
and measurements of $V_{+}(t)$ and $V_{-}(t)$ were performed with an
AD/DA converter AIO-163202F-PE (Contec Co.).
When $V_{\rm{g}}(t)$ was fixed at a constant value between
-6.0~V and -2.45~V, the circuit exhibited limit-cycle
oscillations. Figure~\ref{circuit}(b) shows a limit-cycle orbit on
($V_{+}$, $V_{-}$) plane at $V_{\rm g}(t) \equiv -6.0\;\rm{V}$, and
Figs.~\ref{circuit}(b) and (c) display the corresponding time series
of $V_{+}(t)$ and $V_{-}(t)$.

We repeatedly switched the gate voltage $V_{\rm g}(t)$ between two
values $V_{\rm g1}$ and $V_{\rm g2}$ to simulate a random telegraph signal.
The switching events obeyed a computer-generated Poisson process, namely,
the switching interval $D$ was an exponentially-distributed
random variable with mean interval $\tau = 0.2\;\rm{s}$.
We generated $D$ by the formula $D=-1/\tau\log(1-u)$ with $u$ being computer-generated pseudo random numbers in [0,1].
The typical relaxation time of the circuit to converge to either of the limit cycles was shorter than 0.01~s.
By applying the same time sequence of $V_{\rm g}(t)$ to the circuit
repeatedly, we performed consecutive measurements of the time
series of $V_{+}(t)$, and then repeated this procedure with different
realizations of the random telegraph signals.
In the experiment, $V_{\rm{g1}}$ was fixed to -6.0~V and
  $V_{\rm{g2}}$ was varied between -6.0~V and -2.45~V.

We defined a phase $\phi(t)$ of the circuit state from the time series
of $V_{+}(t)$ as follows. The origin of the phase ($\phi=0$) was taken
as the moment when $V_{+}(t)$ changed its sign from negative to
positive.  Each time $V_{+}(t)$ crossed $0\;\rm{V}$ from negative to
positive, the phase was reset to 0.
Between successive zero-crossing events, the phase was increased with
a constant frequency from 0 to 1.
Note that the frequency of the oscillator was not constant but
changed from cycle to cycle due to random switching of the
driving signal.
From the two consecutive time series of $V_{+}(t)$, we obtained time
series of the absolute phase differences $|\Delta \phi(t)|$ between
two experimental trials (restricted to $[0, 1]$ using the periodicity
of the oscillator) under the same time sequence of $V_{\rm{g}}(t)$.

\subsection{Results}

\begin{figure}[htbp]
  \includegraphics[width=0.8\linewidth]{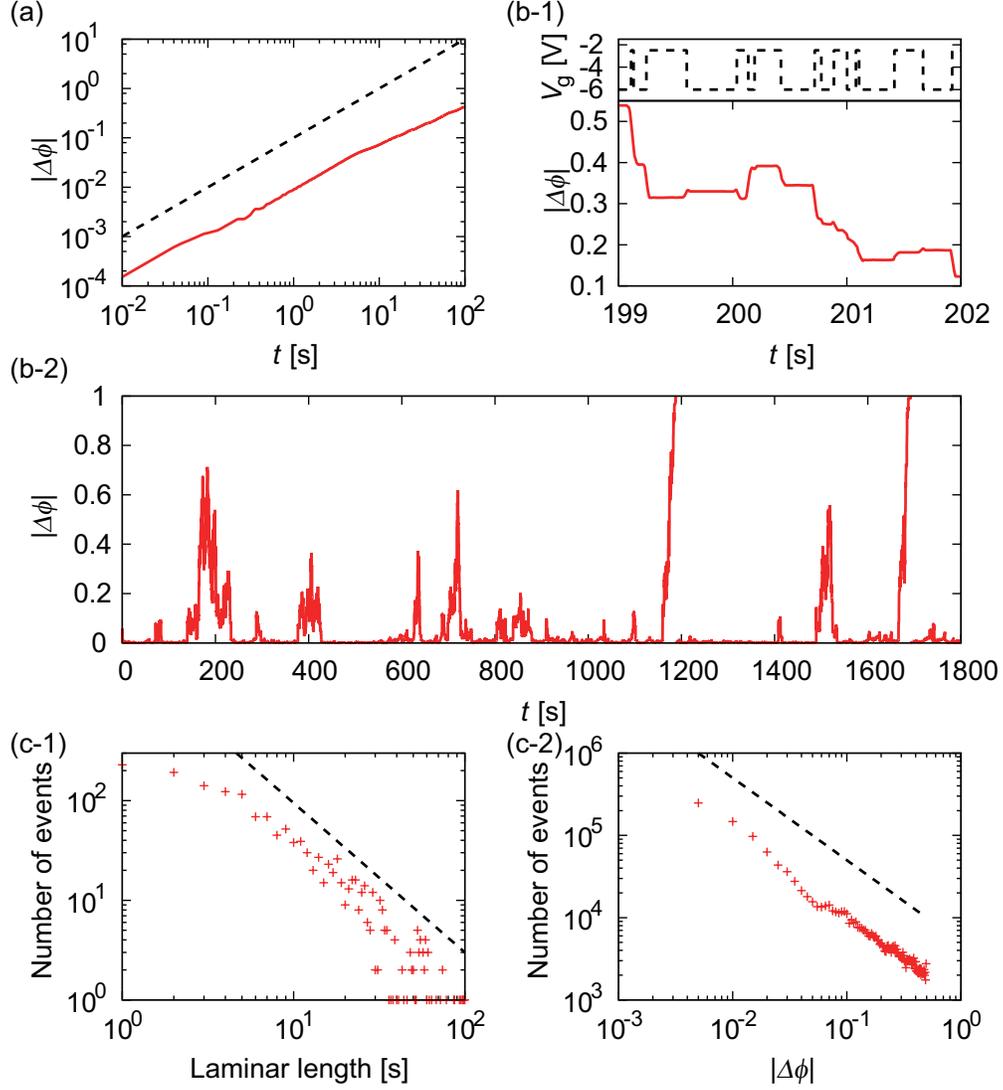}
  \caption{(Color online) (a) Evolution of the absolute phase
    difference $|\Delta\phi(t)|$ under a constant external signal
    $V_{\rm{g}}(t) \equiv -6.0 \;\rm{V}$ depicted using doubly logarithmic
    scales.  The solid curve represents the experimental data and the
    broken line with unit slope represents linear dependence on $t$.
    (b) Time series of the phase difference, $|\Delta\phi(t)|$, with
    $V_{\rm{g}}(t)$ a random telegraph signal ($V_{\rm{g1}} =
    -6.0\;\rm{V}$ and $V_{\rm{g2}} = -2.5\;\rm{V}$).  (b-1) Short time
    series of $V_{\rm{g}}(t)$ (top) and $|\Delta \phi(t)|$
    (bottom). (b-2) Long time series of $|\Delta\phi(t)|$, exhibiting
    noisy on-off intermittency. (c-1) Distribution of the laminar
    length. The broken line represents $t^{-1.5}$. (c-2) Distribution
    of the absolute phase difference $|\Delta\phi(t)|$. The broken
    line represents $\left|\Delta\phi(t)\right|^{-1}$.  }
\label{result}
\end{figure}

Figure~\ref{result}(a) shows the time evolution of the absolute phase
difference $\left| \Delta\phi(t) \right|$ between two experimental
trials under a constant input, $V_{\rm g}(t) \equiv V_{\rm g1} = -
6.0\;\rm V$, which increased linearly with time $t$.
Even after fine tuning, we observe that the average period of oscillations 
differed slightly across experimental trials ($\pm 0.1\;\%$).

Figures~\ref{result}(b-1) and (b-2) show the time series of
$\Delta\phi(t)$ observed under the common telegraph noise,
where $V_{\rm{g}}(t)$ was switched between two values
($V_{\rm{g1}}=-6.0\;\rm{V}$ and $V_{\rm{g2}}=-2.5\;\rm{V}$).
Large changes of $\left| \Delta \phi(t) \right|$ were observed only
in short time windows after the switching
events of $V_{\rm{g}}(t)$ as shown in Fig.~\ref{result}(b-1).
The two experimental trials driven with the same input signal
became mostly synchronized, but
occasionally there were interruptions by short desynchronization events as shown in
Fig.~\ref{result}(b-2).

To characterize this characteristic intermittent behavior of the phase
difference, we measured the distribution of laminar intervals during
which $\left| \Delta \phi(t) \right|$ was smaller than a certain
threshold, $\left| \Delta \phi(t) \right| \leq 0.01$.
As shown in Fig.~\ref{result}(c-1), the stationary distribution of
laminar intervals appeared to follow a power law, whose exponent was
approximately $-1.5$.
We also measured the stationary distribution of the absolute phase
difference $\left| \Delta \phi(t) \right|$, which also exhibited a
power-law tail with an exponent roughly $-1$ (Fig.~\ref{result}(c-2)).
Though the phase difference $\left| \Delta \phi(t) \right|$ shown in
the figure is restricted to the range $[0,1]$, the phase difference
occasionally exhibited jumps of magnitude $1$ due to phase slippage of
one period (e.g. near $t=1200$ and $t=1600$ in
Fig.~\ref{result}(b-2))~\cite{YoshimuraDavisUchida2007}.

Thus, when the gate voltage $V_{\rm g}(t)$ was switched between
two values randomly, different experimental trials tended to be
synchronized even under the effect of slight differences in average
periods and experimental noise.
As we explain later, the characteristic behavior of the phase
difference $\left| \Delta \phi(t) \right|$ was due to noisy on-off
intermittency~\cite{Fujisaka1985,Cenys1997,Pikovsky1992,Nakao1998,NagaiNakaoTsubo2005}.

\section{Analysis}

\subsection{Theory}

Here we briefly summarize our previous theory on the synchronization
of uncoupled oscillators driven by a common random telegraph
signal~\cite{NagaiNakaoTsubo2005}.
Corresponding to the two different values of the input signal
$V_{\rm{g}}(t)$, the circuit exhibits two different limit cycles, LC1
for $V_{\rm{g}}(t) \equiv V_{\rm{g1}}$ and LC2 for $V_{\rm{g}}(t)
\equiv V_{\rm{g2}}$.
When $V_{\rm{g}}(t) \equiv V_{\rm{g1}}$, we can define a phase $0
  \leq \theta_{1}(V_{+}, V_{-}) < 1$ for LC1 and also in the
  $(V_{+}, V_{-})$ plane, which increases with a constant frequency $1
  / T_{1}$ with $T_{1}$ the period of
  LC1~\cite{Kuramoto_book1984,Winfree_book2001}.
Similarly, when $V_{\rm{g}}(t) \equiv V_{\rm{g2}}$, another phase
  $0 \leq\theta_{2}(V_{+}, V_{-}) < 1$ can be defined in the $(V_{+},
  V_{-})$ plane that increases with a constant frequency $1 / T_{2}$,
  where $T_{2}$ is the period of LC2.
The origins of $\theta_{1}$ and $\theta_{2}$ are taken as the points
where $V_{+}(t)$ crosses $0\;\rm{V}$ from negative to positive on LC1 or LC2.
Combining these, we introduce a new phase $\theta(V_{+}, V_{-},
  V_{\rm{g}})$ of the circuit state as
\begin{align}
  \theta(V_{+}, V_{-}, V_{\rm{g}})=\left\{
    \begin{array}{lr}
      \theta_{1}(V_{+}, V_{-}) &(\mbox{when}\; V_{\rm{g}} = V_{\rm{g1}}),
      \\
      \theta_{2}(V_{+}, V_{-}) &(\mbox{when}\; V_{\rm{g}} = V_{\rm{g2}}).
    \end{array}
  \right.
\end{align}

Note that this $\theta$ is different from the phase $\phi$ that
we defined in the previous section by linearly interpolating
successive zero-crossing events.
$\theta$ jumps discontinously at the moments when $V_{\rm{g}}(t)$ switches,
  because $\theta_{1}$ and $\theta_{2}$ increase with strictly
  constant frequencies.
In contrast, $\phi$ is continuous even when $V_{\rm{g}}(t)$ is
fluctuating (except the zero-crossing events of $V_{+}(t)$), but its
frequency differs from cycle to cycle.
When $V_{\rm{g}}(t)$ is kept constant for longer than one period of
  oscillation and than the relaxation time of the circuit to the limit cycle, $\phi$ coincides with $\theta$.
This difference in the definition of the phase variables yields only
small bounded discrepancies in measuring the phase differences between
two time series.

\begin{figure}[htbp]
  \includegraphics[width=\linewidth]{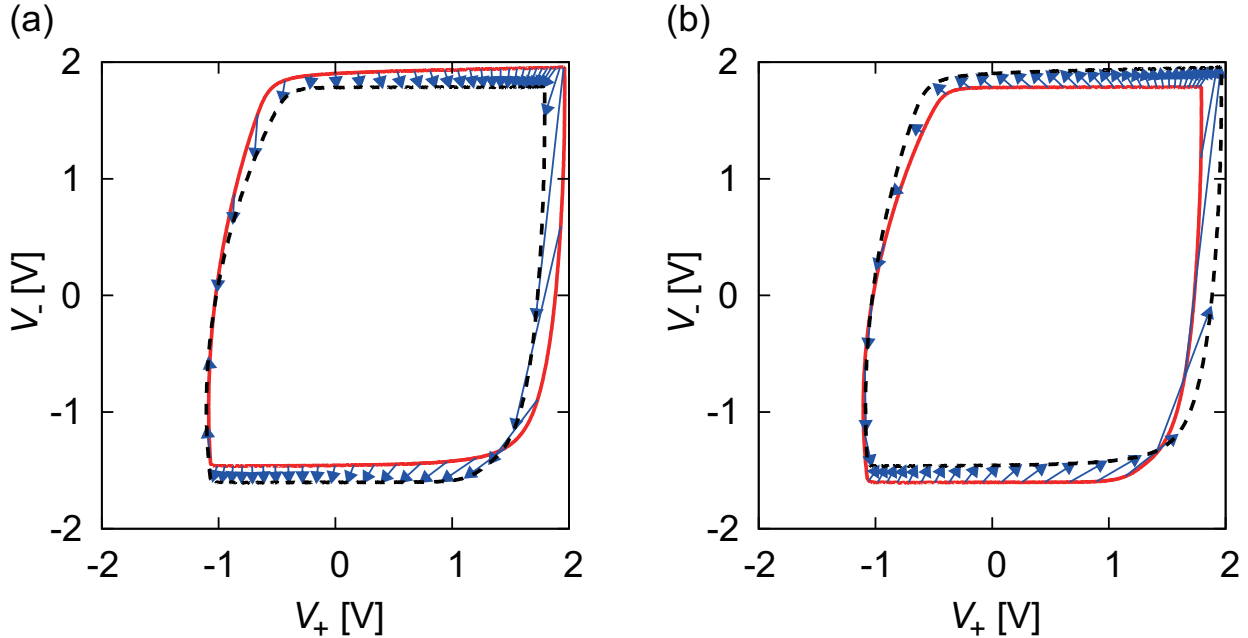}
  \caption{(Color online) Phase mappings between two limit cycles.
    (a) When $V_{\rm{g}}(t)$ switches from $V_{\rm{g1}} =
    -6.0\;\rm{V}$ to $V_{\rm{g2}} = -2.45\;\rm{V}$, points on LC1
    (solid curve) for $V_{\rm{g1}}$ are mapped to the
    corresponding points on LC2 (broken curve) for $V_{\rm{g2}}$ as indicated by arrows.  (b) From $V_{\rm{g2}} =
    -2.45\;V$ to $V_{\rm{g1}} = -6.0\;\rm{V}$.  }
\label{map_schematic}
\end{figure}

We assume that the average switching time of the input signal, $\tau$, is
sufficiently longer than the relaxation time of the orbit to LC1 or
LC2 at fixed $V_{\rm g}(t)$.
In our experiments, we took $\tau = 0.2\ \rm{s}$ and the relaxation
time was typically shorter than 0.01~s, so that this condition was
satisfied.
The orbit of the circuit is then almost always on one of
the limit cycles when $V_{\rm g}(t)$ is switched between two values.
Therefore, we can describe the dynamics of the circuit under randomly
switched $V_{\rm g}(t)$ as alternating phase mappings between
$\theta_{1}$ and $\theta_{2}$ as shown in Fig.~\ref{map_schematic}.
We denote the mapping from $\theta_{1}$ to $\theta_{2}$ that takes
place when $V_{\rm g}(t)$ switches from $V_{\rm g1}$ to
$V_{\rm g2}$ as $\theta_{2} = f_{12}(\theta_{1})$, and the mapping
from $\theta_{2}$ to $\theta_{1}$ when $V_{\rm g}(t)$ switches from
$V_{\rm g2}$ to $V_{\rm g1}$ as $\theta_{1} =
f_{21}(\theta_{2})$~\cite{NagaiNakaoTsubo2005}.

Let us denote the phase on LC1 just before $V_{\rm{g}}(t)$ switches
from $V_{\rm{g1}}$ to $V_{\rm{g2}}$ for the $n$-th time as
$\theta_{1}^{n}$, and the phase on LC2 just before $V_{\rm{g}}(t)$
switches from $V_{\rm{g2}}$ to $V_{\rm{g1}}$ for the $n$-th time as
$\theta_{2}^{n}$.
Then the phase dynamics of the orbit can be described as (assuming
$V_{\rm{g}}(t=0) = V_{\rm{g1}}$)
\begin{align}
  \theta_{1}^{n+1} = f_{21}(\theta_{2}^{n}) + s_{1}^{n+1},\;\;\;
  \theta_{2}^{n} = f_{12}(\theta_{1}^{n}) + s_{2}^{n},
\end{align}
where $s_{1}^{n}$ and $s_{2}^{n}$ are exponentially distributed random
switching intervals whose probability distributions are given by
\begin{align}
  P_{1}(s_{1}^{n})=\frac{T_{1}}{\tau} \exp \left( -
    \frac{s_{1}^{n} T_{1}}{\tau} \right),\;\;
  P_{2}(s_{2}^{n})=\frac{T_{2}}{\tau} \exp \left( -
    \frac{s_{2}^{n} T_{2}}{\tau} \right),
\end{align}
respectively. Small deviations $\Delta \theta_{1}^{n}$, $\Delta
\theta_{2}^{n}$ from $\theta_{1}^{n}$, $\theta_{2}^{n}$ obey
linearized equations
\begin{align}
  \Delta \theta_{1}^{n+1} = f^{\prime}_{21}(\theta_{2}^{n})
  \Delta\theta_{2}^{n},
  \;\;\;
  \Delta \theta_{2}^{n} = f^{\prime}_{12}(\theta_{1}^{n})
  \Delta\theta_{1}^{n},
\end{align}
where $f^{\prime}$ denotes the derivative function
of $f$. After the $n$-th switching,
the amplitude of the deviation $\Delta \theta_{1}^{n}$ is given by
\begin{align}
  |\Delta\theta_{1}^{n}| &= \left( \prod_{i=1}^{n-1} \
    |f^{\prime}_{12}(\theta_{1}^{i})| \
    |f^{\prime}_{21}(\theta_{2}^{i})| \ \right)
  |\Delta\theta_{1}^{1}|,
  \label{dtheta}
\end{align}
and $|\Delta \theta_{2}^{n}|$ similarly.  Thus, for large $n$,
\begin{align}
%%  |\Delta\theta_{1}^{n}| \simeq \exp (\lambda n)
%%  |\Delta\theta_{1}^{1}|
  \frac{|\Delta\theta_{1}^{n}|}{  |\Delta\theta_{1}^{1}|}
  =
  \frac{|\Delta\theta_{2}^{n}|}{  |\Delta\theta_{2}^{1}|}
  \simeq 
  \exp (\lambda n)
\end{align}
holds, where the Lyapunov exponent $\lambda$ is given by
\begin{align}
  \lambda &= \lim_{n \to \infty} \frac{1}{n} \ln \left(
    \prod_{i=1}^{n} \ |f^{\prime}_{12}(\theta_{1}^{i})| \
    |f^{\prime}_{21}(\theta_{2}^{i})| \ \right) = \lambda_{1} +
  \lambda_{2}
\end{align}
with
\begin{align}
  \lambda_{1} \simeq \int^{1}_{0} d\theta_{1} \ln
  \left|f_{12}^{\prime}(\theta_{1}) \right|, \;\;\; \lambda_{2} \simeq
  \int^{1}_{0} d\theta_{2} \ln \left|f_{21}^{\prime}(\theta_{2})
  \right|.
\end{align}
Here, we approximate the average over the stationary distribution of
the phase $\theta_{1}$ or $\theta_{2}$
under the effect of telegraph noises by the average over the
uniform distribution in each equation, because the phases $\theta_{1}$
and $\theta_{2}$ are almost uniformly distributed on LC1 and LC2 when
$\tau$ is sufficiently larger than $T_1$ and $T_2$~\cite{NagaiNakaoTsubo2005}.
In our experiments, we used $\tau = 0.2$, whereas the period of
oscillations was about 0.05~s. Therefore, this condition was
satisfied.

When $\lambda = \lambda_{1} + \lambda_{2} < 0$, phase synchronization
induced by the random telegraph signal is expected. Note that the
switching step $n$ is approximately related to the real time $t$ as $n
\simeq t / 2\tau$, so that
\begin{align}
  \frac{| \Delta \theta (t) |}{| \Delta \theta (0) |} \simeq
  \exp\left( \frac{\lambda}{2 \tau} t \right)
\end{align}
holds for large $n, t$.

\subsection{Determination of phase maps}

We experimentally determined the phase maps $\theta_{1} =
f_{21}(\theta_{2})$ and $\theta_{2} = f_{12}(\theta_{1})$ as follows.
We first measured the period $T_{1}$ of LC1 under a constant input
signal $V_{\rm{g}}(t) \equiv V_{\rm{g1}}$.
After setting $V_{\rm{g}}(t)$ at $V_{\rm{g1}}$ and relaxing the
circuit for 0.5~s, intervals between successive zero-crossing
events of $V_{+}(t)$ from negative to positive values were measured
for 5~s.  $T_{1}$ was determined by averaging these
intervals. We then measured the period $T_{2}$ of LC2 at
$V_{\rm{g}}(t) \equiv V_{\rm{g2}}$ in a similar way.

\begin{figure}
  \includegraphics[width=0.9\linewidth]{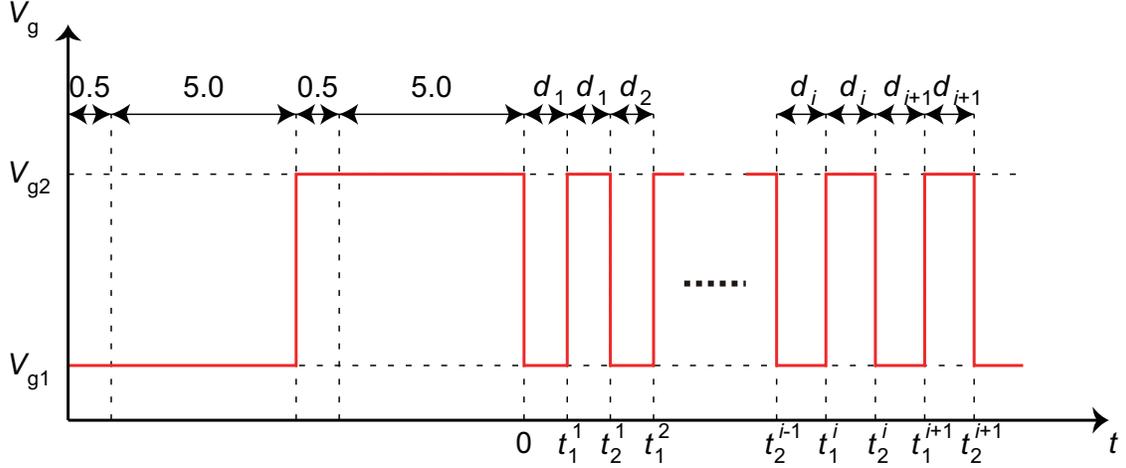}
  \caption{(Color online) The probe signal used to determine the phase
    maps. Before $t=0$, $T_{1}$ and $T_{2}$ were measured.  After $t =
    0$, the phases were measured at $t_{1}^{i}$ and $t_{2}^{i}$ for $1
    \leq i \leq 200$.}
\label{probe_input}
\end{figure}

When the measurement of $T_{2}$ was completed (this moment was defined
as $t=0$), the following {\em regular} telegraph signal (shown
schematically in Fig.~\ref{probe_input}) was applied as the probing
input:
\begin{align}
  V_{\rm{g}}(t) = 
  \left\{
    \begin{array}{ll}
      V_{\rm{g1}}
      &
      \left(
        t_{2}^{i-1} \leq t < t_{1}^{i}
      \right),
      \cr \cr
      V_{\rm{g2}}
      &
      \left(
        t_{1}^{i} \leq t < t_{2}^{i}
      \right),
    \end{array}
  \right.
  \label{map_time_series}
\end{align}
for $i = 1, \cdots, 200$, where the $i$-th switching time from
$V_{\rm{g1}}$ to $V_{\rm{g2}}$ is given by
\begin{align}
  t_{1}^{1} = d_{1}, \;\;\; t_{1}^{i} = d_{i} + \sum_{k=1}^{i-1} 2
  d_{k}\;\;\;(i \geq 2),
\end{align}
and the subsequent $i$-th switching time from $V_{\rm{g2}}$ to
$V_{\rm{g1}}$ is given by
\begin{align}
  %% t_{2}^{0} = 0, \;\;\;
  t_{2}^{i} = \sum_{k=1}^{i} 2 d_{k}\;\;\;(i \geq 1).
\end{align}
Here, $d_{i}$ denotes the length of the $i$-th constant interval of
the input signal. To avoid undesirable synchronization with the probe
signal, we gradually increased $d_{i}$ as $d_{i} = 0.5 + 0.001(i-1)
\;\; (1\leq i\leq200)$.

Using this $V_{\rm{g}}(t)$, we measured the phase $0 \leq
\varPhi_{1}^{i} < 1$ of the circuit at $t_{1}^{i}$,
\begin{align}
  \left\{ ( \varPhi_{1}^{i}, t_{1}^{i} ) \ | \ i = 1 , \cdots, 200
  \right\},
\end{align}
and the phase $0 \leq \varPhi_{2}^{i} < 1$ of the circuit at
$t_{2}^{i}$,
\begin{align}
  \left\{ ( \varPhi_{2}^{i}, t_{2}^{i} ) \ | \ i = 1 , \cdots, 200
  \right\}.
\end{align}
These phases $\varPhi_{1}^{i}$ and $\varPhi_{2}^{i}$ can be identified with
the phases $\theta_1$ and $\theta_2$ used in the theory via
\begin{align}
  \theta_{1}(t=t_{1}^{i}) = \varPhi_{1}^{i}, \;\;\;
  \theta_{2}(t=t_{2}^{i}) = \varPhi_{2}^{i}.
\end{align}
Note that the two types of the phases coincide here because the
  constant intervals $d_i$ of the probe signal are always longer than
  $T_1$, $T_2$, and the relaxation time of the circuit to LC1 or LC2.

At each switching event, the phase jumped from $\varPhi_{1}^{i}$ to
$\tilde{\varPhi}_{2}^{i}$ ($V_{\rm{g1}} \to V_{\rm{g2}}$), or from
$\varPhi_{2}^{i}$ to $\tilde{\varPhi}_{1}^{i}$ ($V_{\rm{g2}} \to
V_{\rm{g1}}$).
The destination phases $\tilde{\varPhi}_{1}^{i}$ and
$\tilde{\varPhi}_{2}^{i}$ just after $t_{1}^{i}$ and $t_{2}^{i}$
were obtained from $\varPhi_{1}^{i}$ and $\varPhi_{2}^{i}$ as
\begin{align}
  \tilde{\varPhi}_{1}^{i} = \left( \varPhi_{1}^{i+1} -
    \frac{d_{i+1}}{T_{1}} \right) \; \mbox{mod}\ 1, \;\;\;
  \tilde{\varPhi}_{2}^{i} = \left( \varPhi_{2}^{i} -
    \frac{d_{i}}{T_{2}} \right) \; \mbox{mod}\ 1,
\end{align}
where the definitions were made modulo $1$ to restrict the phases to $[0,1]$.
Thus, we obtained $200$ realizations of the phase mappings,
\begin{align}
  \left\{ \varPhi_{2}^{i} \to \tilde{\varPhi}_{1}^{i} \ | \ i = 1 , \cdots, 200 \right\},
  \;\;\;
  \left\{ \varPhi_{1}^{i} \to \tilde{\varPhi}_{2}^{i} \ | \ i = 1 , \cdots, 200 \right\}.
\end{align}
We constructed the raw phase maps by piecewise-linearly interpolating
these data as
\begin{align}
  \tilde{\varPhi}_{2} = f_{12}^{\rm{raw}} ( \varPhi_{1} ), \;\;\;
  \tilde{\varPhi}_{1} = f_{21}^{\rm{raw}} ( \varPhi_{2} ),
\end{align}
which were still non-smooth functions due to experimental
fluctuations.

\begin{figure}
  \includegraphics[width=0.7\linewidth]{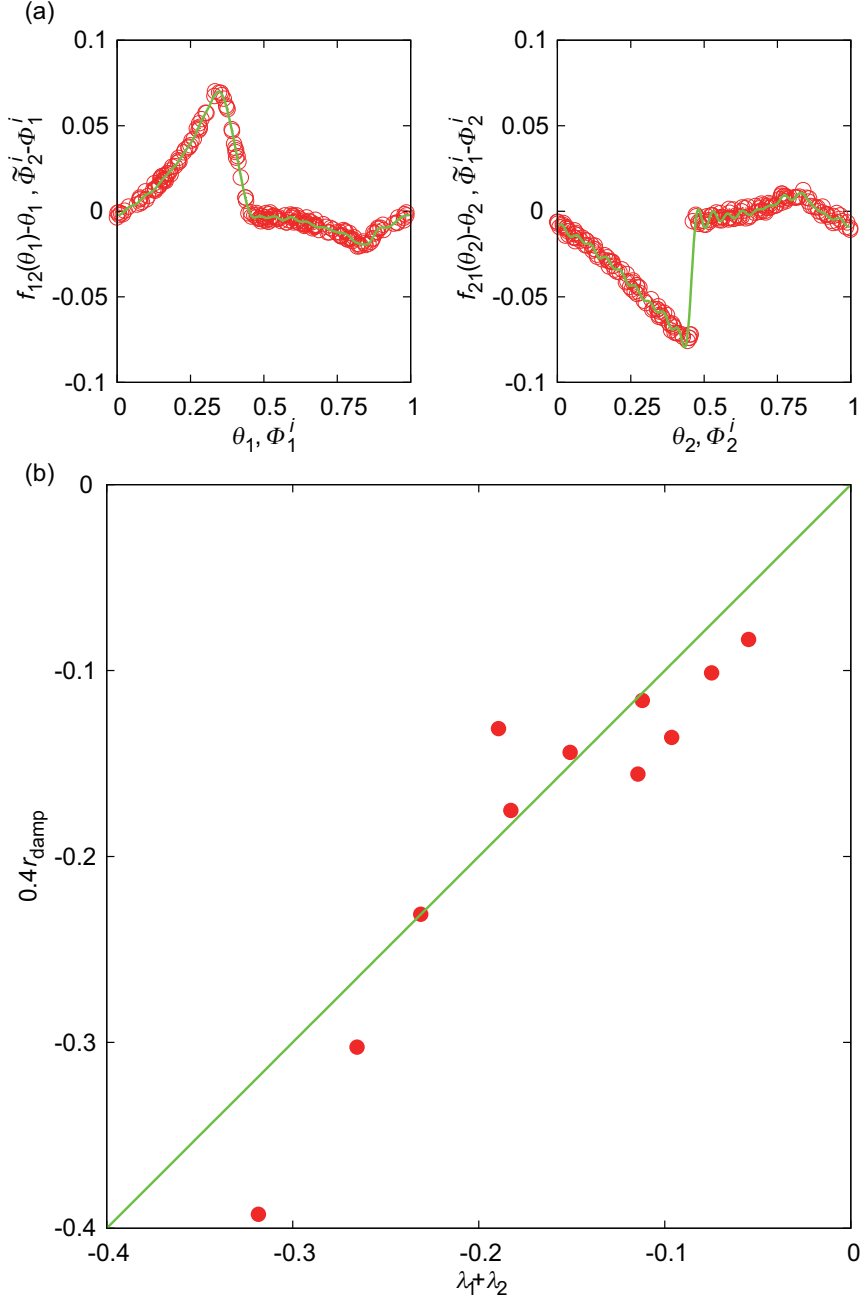}
  \caption{(Color online) (a) Experimentally determined phase
    maps. Circles show the raw data $(\varPhi_{1}^{i},
    \tilde{\varPhi}_{2}^{i}-\varPhi_{1}^{i})$ (left) or
    $(\varPhi_{2}^{i}, \tilde{\varPhi}_{1}^{i}-\varPhi_{2}^{i})$
    (right), and curves represent the non-trivial part of the
    estimated phase maps $f_{12}(\theta_{1}) - \theta_{1}$ (left) or
    $f_{21}(\theta_{2})-\theta_{2}$ (right) obtained after low-pass
    filtering. $V_{\rm{g1}} = -6.0\;\rm{V}$ and $V_{\rm{g2}} =
    -2.5\;\rm{V}$.  (b) Comparison of the Lyapunov exponents $\lambda
    = \lambda_{1} + \lambda_{2}$ with the linear damping rates
    $r_{\rm{damp}}$ directly measured from $|\Delta\phi (t)|$. The
    line represents $0.4\;r_{\rm{damp}} = \lambda_{1} + \lambda_{2}$.
    Filled circles represent the data obtained by varying
    $V_{\rm{g2}}$ ($-2.55\;\rm{V} \leq V_{\rm{g2}} \leq -2.45\;\rm{V}$
    with intervals of $0.01\;\rm{V}$) while fixing $V_{\rm{g1}}$ at
    $-6.0\;\rm{V}$.  }
\label{map}
\end{figure}

Generally, the phase map has a trivial diagonal component, namely, the
identity-map component that exists even when $V_{\rm{g1}} =
V_{\rm{g2}}$, and additional non-trivial components reflecting the
nonlinear transition dynamics between the limit cycles.
We estimated the underlying smooth phase maps $f_{12}$ and $f_{21}$
from the raw phase maps $f_{12}^{\rm{raw}}$ and $f_{21}^{\rm{raw}}$ by
low-pass filtering using the $20$ lowest Fourier modes as
\begin{align}
  f_{12}(\theta_{1}) &= \theta_{1} + \sum_{k=-20}^{20}
  c_{12}^{k} \exp (2\pi k i \theta_{1}), 
  \cr
  f_{21}(\theta_{2}) &= \theta_{2} + \sum_{k=-20}^{20}
  c_{21}^{k} \exp (2\pi k i \theta_{2}),
\end{align}
where $c_{12}^{k}$ and $c_{21}^{k}$ are Fourier coefficients of the
non-trivial components of $f_{12}^{\rm{raw}}$ and $f_{21}^{\rm{raw}}$,
defined as
\begin{align}
  f_{12}^{\rm{raw}}(\varPhi_{1}) - \varPhi_{1}
  &=
  \sum_{k=-\infty}^{\infty} c_{12}^{k}
  \exp (2\pi k i \varPhi_{1}),
  \cr
  f_{21}^{\rm{raw}}(\varPhi_{2}) - \varPhi_{2}
  &=
  \sum_{k=-\infty}^{\infty} c_{21}^{k}
  \exp (2\pi k i \varPhi_{2}).
\end{align}
Figure~\ref{map}(a) displays examples of the phase maps obtained with
the above procedure.

\subsection{Lyapunov exponents and damping rates}

From the experimentally determined phase maps $f_{12}(\theta_{1})$ and
$f_{21}(\theta_{2})$, the Lyapunov exponent $\lambda = \lambda_{1} +
\lambda_{2}$ can be estimated.
This $\lambda$ can be compared with the damping rate $r_{\rm{damp}}$
of small phase differences between two trials subject
to the same telegraph noise.
As we have already explained, the difference in the definition of two
phases results in only a small bounded discrepancy between
$\Delta \theta(t)$ and $\Delta \phi(t)$, so that it does not affect
the Lyapunov exponents or the damping rates.

When the phase difference $\Delta \phi(t)$ is small, we expect the
ensemble average of $\Delta \phi(t)$ over many realizations of the
random telegraph signal to shrink exponentially as
\begin{align}
  \langle | \Delta \phi(t) | \rangle \sim \exp( r_{\rm{damp}} t ),
\end{align}
where $r_{\rm{damp}} < 0$ is the damping rate. As the average
switching interval of $V_{\rm{g}}(t)$ is $\tau$,
\begin{align}
  r_{\rm{damp}} = \frac{\lambda}{2 \tau}
\end{align}
will approximately hold for large $t$ and $n$ provided that the
previous analysis based on the phase-mapping description is
reasonable.

Figure~\ref{map}(b) compares the Lyapunov exponent $\lambda$ with the
damping rate $r_{\rm{damp}}$ obtained for different values of
$V_{\rm{g2}}$ with fixed $V_{\rm{g1}}$.  The damping rate
$r_{\rm{damp}}$ was directly measured from $|\Delta\phi (t)|$ as
\begin{align}
  r_{\rm{damp}} = \frac{1}{M^{-1} \sum_{k=1}^{M} \Delta T_{k}} \ln
  \left| \frac{\Delta \phi_{\rm{lower}}}{\Delta \phi_{\rm{upper}}}
  \right|,
\end{align}
where $\Delta \phi_{\rm{upper}}=0.2$, $\Delta \phi_{\rm{lower}}=0.05$,
$\Delta T_{k}$ is the time needed for $|\Delta\phi|$ to be damped from
$\Delta \phi_{\rm{upper}}$ to $\Delta \phi_{\rm{lower}}$, and $M$ is
the number of such shrinkage events in the time series of $|\Delta
\phi(t)|$.
We measured 20 time sequences of $V_{+}(t)$ for 120 s and calculated
  19 time sequences of the phase difference $\Delta \phi(t)$
  between two consecutive time sequences of $V_{+}(t)$ to obtain
  $r_{\rm{damp}}$.
As shown in Fig.~\ref{map}(c), pairs of $(\lambda, r_{\rm{damp}})$
estimated for various values of $V_{\rm{g2}}$ approximately fall on
the straight line $\lambda = 0.4\;r_{\rm{damp}}$ ($\tau = 0.2$ in the
experiment), which quantitatively verifies the validity of
the phase-mapping description of our experiments.

\subsection{Noisy on-off intermittency}

We have focused so far on the average behavior of the phase
difference.  The phase difference $\left|\Delta \phi(t) \right|$
decreases on average when $\lambda = \lambda_{1}+\lambda_{2} < 0$.
However, as can be seen from Eq.~\eqref{dtheta}, small phase
differences $\Delta \theta_{1}$ and $\Delta \theta_{2}$ are driven
multiplicatively by the random application of two phase maps.
This is a typical situation where noisy on-off (or modulational)
intermittency is expected over long time
scales~\cite{Fujisaka1985,Cenys1997,Pikovsky1992,Nakao1998,
  NagaiNakaoTsubo2005}; due to small noises or heterogeneity inherent
in the system, individual time sequence of $|\Delta \phi(t)|$ can
occasionally grow due to random multiplication even if $\lambda =
\lambda_{1}+\lambda_{2} < 0$, resulting in repetitive transient
bursting.
As already shown in Fig.~\ref{result}(c), this is the case for our
electronic circuit.  The power-law distribution of the laminar
interval with the exponent $-1.5$ as shown in Fig.~\ref{result}(c-1),
and the power-law distribution of the amplitude of the phase
difference as shown in Fig.~\ref{result}(c-2) are consistent with the
theoretical predictions on noisy on-off
intermittency~\cite{Fujisaka1985,Cenys1997,Pikovsky1992,Nakao1998}.

\section{Conclusions}

We investigated synchronization between different experimental
trials induced by common telegraph noises using an electronic
circuit undergoing limit-cycle oscillations.
The dynamics of the circuit could be described in terms of random phase
mappings.
We experimentally determined the phase maps and quantitatively
verified that the Lyapunov exponents determined from the phase maps
agreed with the damping rates measured directly from the time series of
small phase differences.
We also confirmed that noisy on-off intermittency of the phase
difference actually occurs.

The mechanism leading to synchronization that we demonstrated using an
electronic circuit in this paper is general and is expected to be
observed in various systems undergoing limit-cycle oscillations.
%%, for example, in light-sensitive Belousov-Zhabotinsky oscillatory chemical reactions.

\section{Acknowledgments}

We thank Kensuke Arai (Kyoto University, Japan) for helpful
discussions and advice.
This work was supported in part by a Grant-in-Aid for JSPS Fellowships to Ken Nagai (18-3189) and the 21st Century COE (Center for Diversity and Universality in Physics) from the Ministry of Education, Culture, Sports, Science and Technology of Japan.

\end{document}